\documentclass[%
 reprint,
superscriptaddress,
 amsmath,amssymb,
pra,
floatfix,
 array,
]{revtex4-1}

\usepackage{graphicx}% Include figure files
\usepackage{amsthm}
\usepackage{bm}
\begin{document}

%\preprint{APS/123-QED}

\title{Scalable quantum computation architecture using always-on Ising interactions via quantum feedforward}% Force line breaks with \\

\author{Takahiko Satoh}
 \email{satoh@sfc.wide.ad.jp}
\affiliation{
 NTT Basic Research Laboratories, 3-1, Morinosato Wakamiya Atsugi-city,
 Kanagawa 243-0198 Japan}
\affiliation{
 Department of Computer Science, Graduate School of Information
 Science and Technology, The University of Tokyo, 7-3-1, Hongo,
 Bunkyo-ku, Tokyo 113-8656, Japan}
\author{Yuichiro Matsuzaki}
\affiliation{
 NTT Basic Research Laboratories, 3-1, Morinosato Wakamiya Atsugi-city,
 Kanagawa 243-0198 Japan}
\author{Kosuke Kakuyanagi}
\affiliation{
 NTT Basic Research Laboratories, 3-1, Morinosato Wakamiya Atsugi-city,
 Kanagawa 243-0198 Japan}
\author{\mbox{William J. Munro}}
 \affiliation{
 NTT Basic Research Laboratories, 3-1, Morinosato Wakamiya Atsugi-city,
 Kanagawa 243-0198 Japan}
\author{Koichi Semba}
\affiliation{
 Advanced ICT Research Institute, National Institute of Information and
 Communications Technology, 4-2-1, Nukuikitamachi, Koganei-city, Tokyo
 184-8795 Japan}
\author{Hiroshi Yamaguchi}
\affiliation{
 NTT Basic Research Laboratories, 3-1, Morinosato Wakamiya Atsugi-city,
 Kanagawa 243-0198 Japan}
\author{Shiro Saito}
\affiliation{
 NTT Basic Research Laboratories, 3-1, Morinosato Wakamiya Atsugi-city,
 Kanagawa 243-0198 Japan}

\date{\today}% It is always \today, today,

\begin{abstract}
Here, we propose a way to control the interaction between qubits
 with always-on Ising interaction. Unlike the standard method to change
 the interaction strength with unitary operations, we fully make use of
 non-unitary properties of projective measurements so that we can
 effectively turn the interaction on or off via 
 feedforward. Our scheme is useful to generate two- or
 three-dimensional cluster states that are
 universal resources for fault-tolerant quantum computation with this
 scheme, and it provides an alternative
 way to realize a scalable quantum processor.
\end{abstract}

\maketitle

%\tableofcontents

\section{Introduction}
Quantum computation is a new paradigm of information processing. 
Known algorithms give superior performances for tasks such as
factoring~\cite{factoring_1,factoring_2}, searching an unsorted
database~\cite{searching_1,searching_2}, quantum
simulation~\cite{simulation_1,simulation_2}, other algorithms~\cite{qa_dj,qa_linequ,qa_ahs,qa_giso,qa_qwalk,qa_pmatch}  and more.
All these algorithms require a large scale quantum computer.
A quantum computer is composed of a sequence of implementation of
single-qubit gates and two-qubit gates~\cite{univqc_1,univqc_2,univqc_3,univqc_4}.
The single-qubit gate denotes a rotation of the qubit around arbitrary axis and degree. 
A control-phase gate is one of the typical examples of two-qubit
gates. This gate flips the phase of the target-qubit if and only if
the state of the control-qubit is $\lvert 1\rangle$. The roles of
control and target qubit are reversible for control-phase gate.
Individual qubits
should be efficiently addressed and the interaction between two-qubits
should be controlled by some external apparatus. 

The challenge is how to design and build a quantum computer with a
realistic technology. This requires quantum architecture.
There have been a number of these for relevant physical systems,
such as nitrogen-vacancy centre~\cite{Ising_nv,Photonic_NV}, ion
traps~\cite{qc_ion}, and superconducting systems~\cite{Intswitch_sc2}.
Many of those have assumed isolating system and excellent
controllability. However, in realistic circumstances, turning on/off
the interaction in a reliable way is one of the hardest parts in such
architectures. For example, two-qubit gates require in-situ turn on/off the
interaction between qubits by the external control apparatus. Since
imperfection of the interaction control tends to induce correlated
errors between qubits, sophisticated technology is required to suppress
such error rate below the threshold of fault tolerant quantum
computation~\cite{Topological_2D,Topological_2D2,Topological_3D}. However,
varying the interaction between qubits in-situ is 
not possible for all physical systems.

One of the ways to reduce the required level of technology is to use a
system with always-on interaction. There are a couple of theoretical
proposals for this direction. Zhou et al suggested a system with
always-on Heisenberg interaction~\cite{alwaysqc3}. They use interaction free subspace to
protect the target encoded qubit from the residual interaction, and they
show that only local manipulations on the system actually provide
universal quantum computation. Simon et al also suggested to use
always-on Heisenberg interaction system for scalable quantum computation
by collectively tuning the qubits~\cite{alwaysqc,alwaysqc2}. These approach look attractive due to
its simplicity that could reduce potential decoherence from the
interaction. 
 
Here, we propose a novel way to perform universal quantum computation
with a system having an always-on Ising interaction. In quantum
mechanics, there are two type of operations, unitary operations such as
applying microwave pulses and non-unitary operations such as readout of
the qubit. While most of the authors in previous papers use unitary
operation to control the
interaction~\cite{alwaysqc,alwaysqc2,alwaysqc3}, we exploit the
non-unitary 
properties that the projective measurement have. We will assume an
always-on Ising interaction between nearest neighbor qubits, and will
insert an ancillary qubit between the qubits that process quantum
information. We show that it is possible to effectively turn on/off the
interaction via quantum measurement and feedforward on the ancillary
qubits. Since quantum feedforward technology is becoming
matured technology~\cite{qf_1,qf_2,qf_3,qf_4,qf_5,qf_6,qf_7,qf_8,qf_9,qf_10}, our proposal provides a feasible and reliable way to
control the interaction, which is a crucial step for the realization of
quantum information processing.

The remainder of this paper is organized as follows. 
In Sec.~\ref{section_definition}, we review the preliminaries of this
paper.
Section~\ref{section_switching} presents the detail of our scheme to show how always-interaction is effectively
 turned on/off via projective measurement to ancillary qubits and quantum feedforward.
Section~\ref{section_conclusion} concludes our
discussion.

\section{Cluster states as a resource for quantum computation}
\label{section_definition}
A two- or three-dimensional
cluster state can be a universal resource for measurement-based
quantum computation (MBQC)~\cite{oneway,oneway_2,oneway_3,oneway_4} and topological quantum
computation~\cite{Topological_2D,Topological_2D2,Topological_3D}.
A cluster state is composed of $\lvert + \rangle$ state qubits on the
lattice points and controlled-phase gate operation $\hat{U}_{CZ}$
between each pair of nearest-neighbor qubits.
The controlled-phase gate can be realized by Ising type interaction~\cite{oneway,oneway_2}. 
When we consider qubits A and B and Ising type interaction between A and
B, the Hamiltonian to perform controlled-phase gate is as follows
\begin{eqnarray}
\hat{H}_{Ising}=g_{(A,B)}\frac{\bm{1}+\hat{Z}_{A}}{2}\frac{\bm{1}+\hat{Z}_{B}}{2}
\end{eqnarray}
where $g_{\left(A,B\right)}$ denotes the interaction strength between qubits A and B.
By letting a separable state $\lvert ++\rangle _{AB}$ evolve for $g_{(A,B)}t=\pi $ according to this
Hamiltonian, the following unitary operator will be applied to the initial state
\begin{eqnarray}
 \exp\left(-i\pi\frac{\bm{1}+\hat{Z}_{A}}{2}\frac{\bm{1}+\hat{Z}_{B}}{2}\right)
=U_{CZ}^{\left(A,B\right)}.
\end{eqnarray}
and hence we can create the controlled-phase gate.

Although there are many proposal to realize Ising type
interaction such as ultracold atoms in an optical
lattice~\cite{Ising_ol,Ising_ota,Ising_ol1,Ising_ol11,Ising_ol12,Ising_ol2,Ising_ol21,Ising_ol3},
ion traps~\cite{Ising_ti1,Ising_ti2,Ising_ti21,Ising_ti22,Ising_ti3,Ising_ti4}, superconducting charge 
qubit~\cite{Ising_sc_charge}, superconducting spin
qubit~\cite{Ising_scspin}, superconducting flux qubit~\cite{Ising_flux}, 
resonator waveguide~\cite{Ising_wg}, nitrogen-vacancy
center~\cite{Ising_nv00,Ising_nv0,Ising_nv1,Ising_nv,Ising_nv11,Ising_nv12,Ising_nv2},
quantum dot~\cite{Ising_qd0,Ising_qd,Ising_qd1,Ising_qd2} and
electronic spins coupled to the motion of magnetized mechanical resonators~\cite{Ising_spin},
the major
challenge for experimental realization is to switch on/ off the
interaction with a high fidelity. Only a few experiments have
demonstrated a high fidelity controllable two-qubit gate with a fidelity
above the threshold of fault tolerant quantum computation~\cite{highfgate_ion,highfgate_ion2,highfgate_sc}.
One of the possible ways to overcome the experimental
difficulties for demonstrating the high-fidelity two-qubit gates is to
use an always-on interaction scheme~\cite{always_switching,always_switching2,alwaysqc4,alwaysqc,alwaysqc2,alwaysqc3}.
 Since there are no needs for the
additional controlling operations to switch the interaction, these
scheme may scale well for a large number of qubits. Here, we propose a
new approach to implement the controlled-phase gate tolerant quantum computation with
always-on interaction by using the non-unitary properties of projective
operations and quantum feedforward.

\section{Effective interaction control via projective measurements and quantum feedforward}
\label{section_switching}
\subsection{Effective turn on/off interaction by measurement and
  quantum feedforward}
\label{switchingwithqf}
We introduce the Hamiltonian to realize our scheme to
turn on/off the interaction effectively via projective measurements and
quantum feedforward. The physical device that we consider is a general solid-state
system where every qubit can be individually controlled by a microwave
pulse and there are always-on interactions between nearest neighbor
qubits.
We assume the following two qubit Hamiltonian.
\begin{eqnarray}
 \hat{H}_{AB}&=&\hat{H}_{local}+ \hat{H}_{interaction} \\
\hat{H}_{local}&=&\!\sum^{}_{j=A,B}\!\left(\frac{\omega_j}{2}\hat{Z}_{j} +\! \lambda_{j}(t)
 \cos{\!\left(\omega'_{j}t +\! \theta\right)}\hat{X}_{j}\right)\\
\hat{H}_{interaction}&=&\frac{g_{\left(A,B\right)}}{4}\hat{Z}_{A}\hat{Z}_{B}
\end{eqnarray}
where $\omega$, $\lambda(t)$, $\omega'$, $\theta$ and $g$ denote the qubit energy, Rabi
frequency, microwave frequency, a phase of the microwave, and
interaction strength.
In most of the solid-state systems, it is  possible to
control the value of $\lambda(t)$ by changing the power of
microwave with much higher accuracy than the case of two-qubit gates.
We move to a rotating frame defined by
\begin{eqnarray}
 \hat{U}_{AB}=\exp\left(-i\sum_{j=A,B}\frac{\omega'_j}{2}
\hat{Z}_jt\right)
\end{eqnarray}
 where $\omega'_{j}$ denotes its angular frequency of the rotating frame at
the site $j$, and use a rotating wave
approximation so that we could obtain the following Hamiltonian
\begin{eqnarray}
\label{hamiltonianrwa}
\hat{H}_{AB}&\simeq&\!\!
\sum^{}_{j= A,B}\!\!\left(\frac{\omega_{j}\!-\!\omega'_{j}}{2}\hat{Z_j}\!+\!\frac{\lambda_{j}(t)}{2}\hat{A}^{\theta}_{j}\right)
\!+\!\frac{g_{\left(A,B\right)}}{4}\hat{Z}_{A}\hat{Z}_{B}
\end{eqnarray}
where 
\begin{eqnarray}
\hat{A}^{\theta}=\begin{pmatrix}
 0&e^{-i\theta}\\
e^{i\theta}&0 \end{pmatrix}.
\end{eqnarray}
 Unless when required to
perform single qubit gates, we turn off the microwave and set
$\lambda =0$, and therefore the Hamiltonian introduced here is
effectively the same as an
Ising model with always-on interaction. On the other hand, for the
implementation of accurate single-qubit rotations, we assume a large
Rabi frequency as $\lambda \gg g$ so that the coupling strength from the
nearest neighbor qubit can be negligible.

The Hamiltonian described above has an interesting
property that an interaction from the other qubit can
be turned off by preparing the state of a qubit in a ground state.
To explain this, we choose the microwave frequencies
  as follows
\begin{eqnarray}
 \omega'_A = \omega_A - \frac{1}{2}g_{(A,B)},
 \omega'_B = \omega_B - \frac{1}{2}g_{(A,B)},
\end{eqnarray}
 and set
\begin{eqnarray}
\lambda_{A}=\lambda_{B}=0.
\end{eqnarray}
Then, the Hamiltonian (\ref{hamiltonianrwa}) becomes as follows.
\begin{eqnarray}
\hat{H}'_{AB} &=& \sum_{j= A,B}\frac{\omega_j -
\omega'_{j}}{2}\hat{Z}_{j}+\frac{g_{(A,B)}}{4}\hat{Z}_{A}\hat{Z}_{B}\\
&=& \sum_{j= A,B}\frac{g_{(A,B)}}{4}\hat{Z}_{j} +\frac{g_{(A,B)}}{4}\hat{Z}_{A}\hat{Z}_{B}\\
&=& g_{(A,B)}\frac{\mathbf{1}+\hat{Z}_{A}}{2}\frac{\mathbf{1}+
  \hat{Z}_{B}}{2}\label{hamiltoniantotal}
\end{eqnarray}

Interestingly, if the qubit A is prepared in a ground state, the
interaction from the qubit A  cancels out because of
\begin{eqnarray}
g_{(A,B)}\frac{\mathbf{1}+\hat{Z}_{A}}{2}\frac{\mathbf{1}+\hat{Z}_{B}}{2}\lvert\downarrow\rangle
 _A=0.
\end{eqnarray}
This means that preparing a specific qubit in a ground state
effectively turn off the interaction between this qubit and
nearest-neighbor qubit.
Therefore, if all nearest-neighbor qubits are ground state, the
 qubit is not affected by any interactions,
 which is the striking feature of our scheme.
Also,  if the qubit A is prepared in a excited state, the interaction
cause the extra phase shift to the qubit B.

It is worth mentioning that we need a precise control of the frequency
of the microwave in our scheme.
We investigate the effect of a small detuning from the target
frequency of the microwave. Suppose that there is a detuning of $\delta
\omega _j$ from the the target frequency, we have
\begin{eqnarray}
\omega'_{j}=\omega_{j}-\frac{g_{(A,B)}}{2}+\delta \omega'_{j}.
\end{eqnarray}
In this case, we can rewrite the
Hamiltonian (\ref{hamiltoniantotal}) as follows.
\begin{eqnarray}
\hat{H}''_{AB} =g_{(A,B)}
\frac{\mathbf{1} + \hat{Z}_{A}}{2}\frac{\mathbf{1} + \hat{Z}_{B}}{2} 
-\sum_{j=A,B}\frac{\delta\omega'_j}{2}\hat{Z}_{j}.
\end{eqnarray}
Hence, frequency errors cause phase shift error on each qubit.
Fortunately, due to recent development of the microwave technology, an
accurate control of the microwave frequency is available. Therefore,
in this paper, we assume that we can choose the exact microwave
frequency to avoid this kind of error.

\subsection{Implementation of controlled-phase gate}
\begin{figure}[htbp]
\includegraphics[width=86mm]{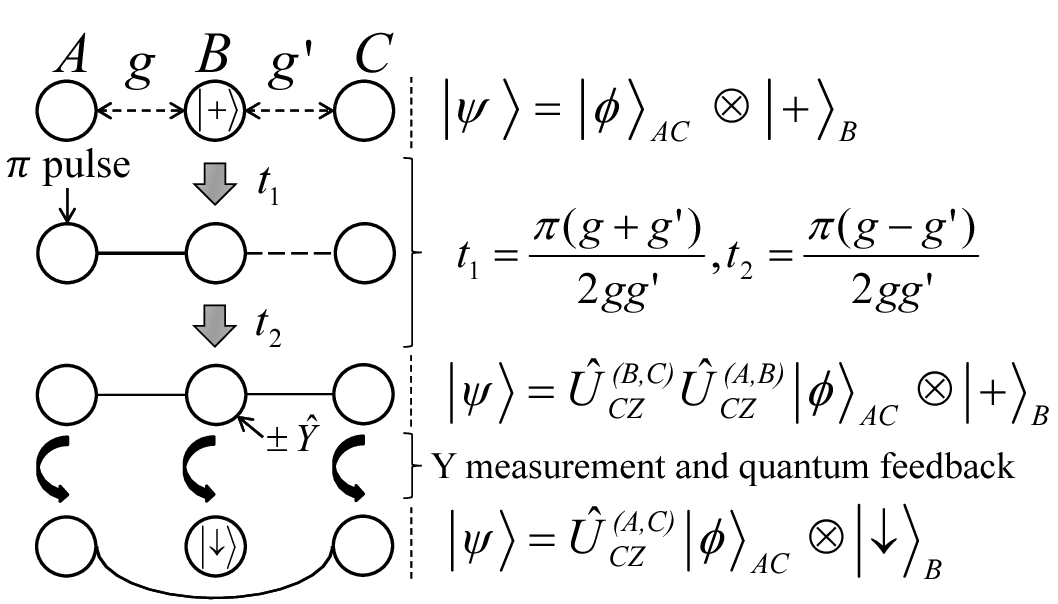}
\caption{\label{echopulse} Schematic of our scheme to implement two-qubit gates via projective
 measurements and quantum feedforward under the effect of always-on
 Ising interaction. First, we let evolve the state $|\phi \rangle
 _{AC}\otimes |+\rangle _{B}$ for a time $t_{1}$ according to the Hamiltonian.
 Second, we perform $\pi$ pulse on the middle qubit B.
 Third, we let evolve the state for a time $t_{2}$.
 Finally, we perform a projective measurement and quantum feedforward
 on the qubit B, so that a
 controlled-phase gate can be implemented  between qubits A and C. 
 Due to the engineered Hamiltonian form that we make, the interaction
 between qubits is turned off as long as the qubit B is in a ground state.
 }
\end{figure}

\begin{figure}[htbp]
\includegraphics[width=86mm]{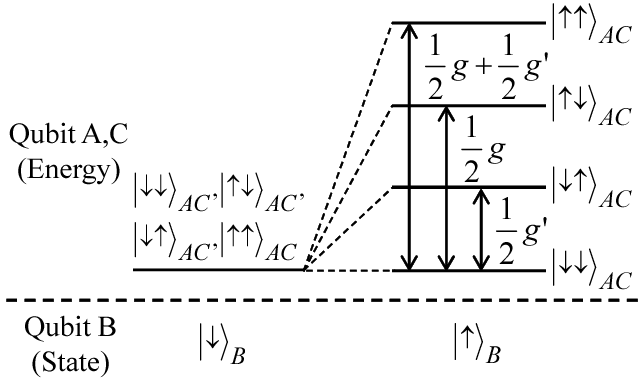}
\caption{\label{diagram}
The energy diagrams of qubit A and C. The energies depend on the state
of the qubit B. The energies of qubit A and C are degenerate when
the qubit B is in a ground state. However, once the qubit B is
excited, degeneracy is removed so that the energy difference occurs
between the states of qubit A and C.}
\end{figure}

We start to illustrate our concept about how to
control the effective interaction via projective measurements and
quantum feedforward.
Suppose that we have three qubits A, B, and C in a raw, and the
coupling strengths between the nearest neighbor qubits are $g_{(A,B)}$
and $g'_{(B,C)}$ as shown in Fig.~\ref{echopulse} where we assume $g>g'$
without loss
of generality.
Then, the system Hamiltonian becomes as follows.
\begin{eqnarray}
\label{Hamiltonian_3qubit}
 \hat{H}&=&\sum^{}_{j=A,B,C}\left(\frac{\omega_j}{2}\hat{Z}_{j} + \lambda_{j}(t)
 \cos{\left(\omega'_{j}t + \theta\right)}\hat{X}_{j}\right)\nonumber\\
&&+\frac{g_{\left(A,B\right)}}{4}\hat{Z}_{A}\hat{Z}_{B}+\frac{g'_{\left(B,C\right)}}{4}\hat{Z}_{B}\hat{Z}_{C}\\
&\simeq&
g_{(A,B)}\frac{\mathbf{1}\!+\!\hat{Z}_{A}}{2}\frac{\mathbf{1}\!+\!\hat{Z}_{B}}{2}
+g'_{(B,C)}\frac{\mathbf{1}\!+\!\hat{Z}_{B}}{2}\frac{\mathbf{1}\!+\!\hat{Z}_{C}}{2}
\end{eqnarray}
with
\begin{eqnarray}
\omega'_A &=& \omega_A - \frac{1}{2}g_{(A,B)},\omega'_B =
\omega_B - \frac{g^{}_{(A,B)}+g'_{(B,C)}}{2},\nonumber\\
\omega'_C &=& \omega_C - \frac{1}{2}g'_{(B,C)},\lambda_{A}=\lambda_{B}=\lambda_{C}=0.
\end{eqnarray}
As written in Eq.~(\ref{Hamiltonian_3qubit}), the state of the qubit B
changes the energies of qubits A and C.  
When we set the qubit B to ground state, all eigen states of qubits A and C
degenerate therefore $\hat{H}$ does not
change the system in time.
We show those energy diagrams in Fig.~\ref{diagram}.

The ancillary qubit induces a conditioned dynamics.  The excited state
of the ancillary qubit causes the phase rotation on the other qubits,
while the ground state of the ancillary qubit does not induce any
phase
shift on them. Therefore, if we have a superposition of the ancillary
qubit, the other two qubits are entangled via such a conditioned
dynamics. In order to see this effect more clearly, we describe how such conditioned dynamics
occur in Appendix~\ref{details_scheme}.

Here, we show the procedure of our scheme for
  controlled-phase gate.
Firstly, we prepare a separable $|+\rangle $ state for
the qubit B, and prepare an arbitrary state for qubits A and C.
An initial state is described by
\begin{align}
\rho=\rho_{AC}\otimes\lvert + \rangle \langle + \rvert_{B}.
\end{align} 

Secondly, we let the state evolve for a time $t_1$,  perform $\pi$ pulse to
the qubit $A$, and let the state evolve for a time $t_2$.
Here, we adopt a spin echo technique
\cite{Compositepulse,Compositepulse2,Compositepulse3} to balance the
interaction. 
In the spin echo technique, implementation of a  $\pi$
pulse can refocus the dynamics of the spin so that the effect of the
interaction should be cancelled out.
We therefore introduce 
\begin{eqnarray}
\label{time_one}
t_1=\frac{\pi\left(g_{(A,B)}+g'_{(B,C)}\right)}{2g_{(A,B)}g'_{(B,C)}}
\end{eqnarray}
 and 
\begin{eqnarray}
\label{time_two}
t_2=\frac{\pi\left(g_{(A,B)}-g'_{(B,C)}\right)}{2g_{(A,B)}g'_{(B,C)}}
\end{eqnarray}
 to satisfy
\begin{eqnarray}
g_{(A,B)}(t_1-t_2)=g'_{(B,C)}(t_1+t_2)=\pi.
\end{eqnarray}
The total unitary evolution $\hat{U}_{CZ}^{(A,B)}\hat{U}_{CZ}^{(B,C)}$
can be described by
\begin{eqnarray}
\hat{U}=\mbox{exp}\Big(&-&ig_{(A,B)}(t_1-t_2)\frac{\bm{1}+\hat{Z}_{A}}{2}\frac{\bm{1}+\hat{Z}_{B}}{2}\nonumber\\
&-&ig'_{(B,C)}(t_1+t_2)\frac{\bm{1}+\hat{Z}_{B}}{2}\frac{\bm{1}+\hat{Z}_{C}}{2}\Big),
\end{eqnarray}
up to  local equivalent,
 so that we can perform
controlled-phase gates even if the coupling strength is
asymmetric. The details are explained in Appendix~\ref{details_scheme}.

Thirdly, we perform $\hat{Y}$ basis 
\begin{eqnarray}
\lvert \pm 1_{\hat{Y}} \rangle =
\frac{1}{\sqrt{2}}(\lvert \uparrow \rangle \pm i\lvert \downarrow \rangle )
\end{eqnarray}
 measurement
on the middle qubit B. The state after the measurement is written as 
\begin{eqnarray}
\rho'_{\pm}
 =\hat{P}_{B}^{\pm}e^{-i\hat{H}t}\rho e^{i\hat{H}t}\hat{P}_{B}^{\pm}
\end{eqnarray}
 where $\pm $ denotes the
 measurement result. 
Here, 
\begin{eqnarray}
\hat{P}^{\pm}= \frac{1}{2}\left(\mathbf{1}\pm\hat{Y}\right)
\end{eqnarray}
denotes a projection operator on the qubit B.
Finally, we perform a quantum feedforward
operation, that is an implementation of
different local operations depending on the measurement results, onto
the qubit B so that the qubit B can be prepared in a ground state.
We define a feedforward operator as
\begin{eqnarray}
\hat{F}^{\pm}_{ABC}=\hat{S}^{\pm}_{A}\hat{U}^{\frac{\mp\pi}{2},\hat{X}}_{B}\hat{S}^{\pm}_{C}
\end{eqnarray}
 where $\hat{S}^{\pm}$ denotes a shift gate defined as 
\begin{eqnarray}
 \hat{S}^{\pm}=\begin{pmatrix}1&0\\0&\pm i\end{pmatrix}
\end{eqnarray} 
and $\hat{U}^{\theta,\hat{X}}$
 denotes a single-qubit rotating around
   $x$-axis rotation with an angle of $\theta $.
The state after the quantum feedforward is described as
\begin{eqnarray}
\rho_{\text{final}}&=&\hat{F}_{ABC}^{+}\rho'_{+}
\hat{F}^{+\dag}_{ABC}+\hat{F}_{ABC}^{-}\rho'_{-}
\hat{F}^{-\dag}_{ABC}\\
&=&\hat{U}_{CZ}^{(A,C)}\rho_{AC}
\hat{U}_{CZ}^{(A,C)}\otimes \lvert \downarrow \rangle \langle
\downarrow\rvert_{B}.
\end{eqnarray}
Therefore, after these operations, controlled-phase operations are
performed between qubits A and C, and the state does not evolve anymore
because the qubit B is prepared in a ground state.
As shown in Fig.~\ref{diagram}, the states of qubits A and C degenerate and
hence interactions are effectively turned off.

Meanwhile, if we set the qubit B in an excited state by
quantum feedforward operation, the final state become as follows.
\begin{eqnarray}
\rho'_{\text{final}}&=&e^{-i\hat{H}t'}\!\left(\hat{U}_{CZ}^{(A,C)}
\rho_{AC} \hat{U}_{CZ}^{(A,C)}\otimes \lvert \uparrow \rangle \langle
\uparrow\rvert_{B}\!\right)\!e^{i\hat{H}t'}\\
&=&e^{-i\hat{H'}t'}\hat{U}_{CZ}^{(A,C)}
\rho_{AC} \hat{U}_{CZ}^{(A,C)}e^{i\hat{H'}t'}\otimes \lvert \uparrow \rangle \langle\uparrow\rvert_{B}
\end{eqnarray}
where $\hat{H'}$ denotes the following Hamiltonian
\begin{eqnarray}
\label{Hamiltonian_H'}
\hat{H'}=g_{(A,B)}\frac{\mathbf{1}+\hat{Z}_{A}}{2}+g'_{(B,C)}\frac{\mathbf{1}+\hat{Z}_{C}}{2}.
\end{eqnarray}
The energy eigenstates are not degenerate as shown in
Fig.~\ref{diagram} and hence interactions cause the extra phase shift to qubits A and C.
In principle, we can correct these extra phases by performing single
qubit rotation. However, unless single qubit rotation can be perfectly
performed, such operations induce another error, which makes it
difficult to perform fault-tolerant quantum computation.
In addition, it is usually difficult to keep the state in an exited
state due to the existent of the energy relaxation.
For these reasons, we set
the qubit B in a ground state after the projective measurement.

It is worth mentioning that, although we introduce a three-qubit case
as an example, it is straightforward to generalize this idea into a
multi-qubit case to create a two or three dimensional cluster state. 

Since the interaction is
Ising type, the eigenvectors are represented by the computational
basis ($\lvert \uparrow \rangle, \lvert \downarrow \rangle$ basis). This means that the ancillary qubit induces a conditional
dynamics such that the target qubits evolve differently depending on
the state of the ancillary qubit. If we have a superposition of the
ground state and excited state of the ancillary qubit, it becomes
possible to realize the superposition of such two dynamics. This is
the key to entangle the ancillary qubit with the target qubits.

\section{Conclusion}
\label{section_conclusion}
Here we show a way to perform controlled-phase gate operation 
  with always-on Ising interaction. Our method uses projective measurements and quantum
  feedforward to effectively turn the interaction on or off in this
  system. Importantly, a direct control of the interaction is not
  required in our scheme. Therefore, our protocol  would provide a
  practical way to implement two-qubit gates for a system where an
  interaction is always-on, which is an important step for scalable
  quantum computation.

\begin{acknowledgements}
We would like to thank Keisuke Fujii for useful discussions.
\end{acknowledgements}

\appendix
\section{The details of implementation of controlled-phase gate}
\label{details_scheme}
In this appendix, we show the details of our scheme
to perform controlled-phase gate.
We set the initial state of the system as following.
\begin{eqnarray}
\lvert\Psi_{1}\rangle&=&\lvert +\downarrow+\rangle_{ABC}.
\end{eqnarray}
From the Hamiltonian $\hat{H}'$ in Eq.~(\ref{Hamiltonian_H'}), we define the
following local Hamiltonians.
\begin{eqnarray}
\hat{H}'_A=g_{(A,B)}\frac{\mathbf{1}+\hat{Z}_{A}}{2},\hat{H}'_C=g'_{(B,C)}\frac{\mathbf{1}+\hat{Z}_{C}}{2}.
\end{eqnarray}

Firstly, we perform $\frac{\pi}{2}$ pulse to the qubit B,
so that we can obtain the following state
\begin{eqnarray}
\lvert\Psi_{2}\rangle&=&\lvert +++\rangle_{ABC}.
\end{eqnarray} 

Secondly, we let the state evolve for a time $t_{1}$ in Eq.~(\ref{time_one}).
The state becomes as follows.
\begin{eqnarray}
\lvert\Psi_{3}\rangle
&=&\frac{1}{\sqrt{2}} (\lvert \downarrow \rangle_{B} 
+ e^{i(\hat{H}'_{A}+\hat{H}'_{C})t_{1}}\lvert \uparrow \rangle_{B})\otimes \lvert ++\rangle_{AC}
\end{eqnarray} 

Thirdly, we perform $\pi$ pulse to the qubit A to balance the effects of
interaction strengths. 
\begin{eqnarray}
\lvert\Psi_{4}\rangle
&=&  \frac{1}{\sqrt{2}}( \lvert \downarrow \rangle_{B} \lvert
+ +\rangle_{AC}\nonumber\\
&&+\hat{X}_{A}e^{i(\hat{H}'_{A}+\hat{H}'_{C})t_{1}} \lvert
\uparrow \rangle_{B} \lvert ++\rangle_{AC})
\end{eqnarray} 

Fourthly, we let the state evolve for a time $t_{2}$ in Eq.~(\ref{time_two}).
Controlled-phase gates are performed between two pairs of qubits as follows.
\begin{eqnarray}
\lvert\Psi_{5}\rangle
&=&\frac{1}{\sqrt{2}}( \lvert \downarrow \rangle_{B} \lvert
++\rangle_{AC} \nonumber\\
&&+ e^{i(\hat{H}'_{\!A}+\hat{H}'_{\!C})t_{2}}\hat{X}_{\!A}
e^{i(\hat{H}'_{\!A}+\hat{H}'_{\!C})t_{1}} \lvert 
\uparrow  \rangle_{\!B} \lvert ++\rangle_{\!AC})\\
&=&\frac{1}{\sqrt{2}}( \lvert \downarrow \rangle_{B} \lvert
++\rangle_{AC} \nonumber\\
&& + \hat{X}_{A} e^{i\hat{H}'_{A}(t_{1}-t_{2})}
 e^{i\hat{H}'_{C}(t_{1}+t_{2})} \lvert 
\uparrow \rangle_{B} \lvert ++\rangle_{AC})\\
\label{Psi_5_final}
&=&\frac{1}{\sqrt{2}}( \lvert \downarrow \rangle_{B} \lvert
++\rangle_{AC} -  \lvert \uparrow \rangle_{B} \lvert --\rangle_{AC}).
\end{eqnarray} 

Finally, we measure the qubit B on
Y-basis. According to the measurement result, the states becomes as follows.
\begin{eqnarray}
\lvert\Psi^{\pm}_{6}\rangle
&=&\frac{1}{2}((\lvert \uparrow \rangle_{B}\pm i\lvert
\downarrow \rangle_{B}) \lvert ++\rangle_{AC}\nonumber\\
&&  \mp i(\lvert \uparrow \rangle_{B} \pm i \lvert \downarrow
\rangle_{B})\lvert --\rangle_{AC})\\
&=&\frac{1}{\sqrt{2}}\lvert \pm 1_{\hat{Y}} \rangle_{B} \otimes (\lvert ++\rangle_{AC}\mp i\lvert --\rangle_{AC}).\\
\end{eqnarray} 

The operation of quantum feedforward is determined according to the
measurement result. These operations are equivalent to perform a controlled-phase gate
between qubits A and C as follows. 
\begin{eqnarray}
\lvert\Psi^{\pm}_{7}\rangle
&=& \hat{F}_{ABC}^{\pm}\lvert\Psi^{\pm}_{6}\rangle\\
&=& \frac{1}{\sqrt{2}} \hat{U}^{\frac{\mp\pi}{2},\hat{X}}_{B}
\lvert \pm 1_{\hat{Y}} \rangle_{B}\nonumber\\
&&\otimes \hat{S}^{\pm}_{A}\hat{S}^{\pm}_{C}(\lvert
++\rangle_{AC}  \mp i\lvert --\rangle_{AC})\\
&=& \frac{1}{\sqrt 2}(\lvert \uparrow -\rangle_{AC} - \lvert
\downarrow + \rangle_{AC})\otimes\lvert \downarrow\rangle_{B}\\
&=& \hat{U}^{(A,C)}_{CZ}\lvert\Psi_{1}\rangle.
\end{eqnarray} 

\bibliography{201312_PRA_fixed}% Produces the bibliography via BibTeX.

\end{document}